\documentclass[sigconf, nonacm]{acmart}
\usepackage{booktabs}
 \usepackage[normalem]{ulem}
\useunder{\uline}{\ul}{}

\usepackage{graphicx}
\usepackage{subcaption}
\usepackage{multirow}

\usepackage{amsmath}
\usepackage{amssymb}
\usepackage{amsthm}
\usepackage{mathtools}
\usepackage{bbm}
\def\R{\mathbb{R}}

\newcommand\figref{\textbf{Figure}~\ref}
\newcommand\tabref{\textbf{Table}~\ref}
\newcommand\equref{\textbf{Equation}~\ref}

\AtBeginDocument{%
  \providecommand\BibTeX{{%
    \normalfont B\kern-0.5em{\scshape i\kern-0.25em b}\kern-0.8em\TeX}}}

\setcopyright{acmcopyright}
\copyrightyear{2018}
\acmYear{2018}
\acmDOI{10.1145/1122445.1122456}

\acmConference[Woodstock '18]{aaa}{aaa}{aaa}
\acmBooktitle{Woodstock '18: ACM Symposium on Neural Gaze Detection,
  June 03--05, 2018, Woodstock, NY}
\acmPrice{15.00}
\acmISBN{978-1-4503-XXXX-X/18/06}

\begin{document}

\title{Learning to Recommend Signal Plans Under Incidents With Real-time Traffic Prediction}
\titlenote{To appear in Transportation Research Record: Journal of the Transportation Research Board: \url{https://doi.org/10.1177/0361198120917668}}
\author{Weiran Yao}
\affiliation{%
  \institution{Carnegie Mellon University}
  \streetaddress{5000 Forbes Avenue}
  \city{Pittsburgh}
  \state{PA}
  \country{USA}}
\email{weiran@cmu.edu}

\author{Sean Qian}
\authornote{Corresponding author}
\affiliation{%
  \institution{Carnegie Mellon University}
  \streetaddress{5000 Forbes Avenue}
  \city{Pittsburgh}
  \state{PA}
  \country{USA}}
\email{seanqian@cmu.edu}

\renewcommand{\shortauthors}{Yao and Qian}

\begin{abstract}
 The main question to address in this paper is to recommend optimal signal timing plans in real time under incidents by incorporating domain knowledge developed with the traffic signal timing plans tuned for possible incidents, and learning from historical data of both traffic and implemented signals timing. The effectiveness of traffic incident management is often limited by the late response time and excessive workload of traffic operators. This paper proposes a novel decision-making framework that learns from both data and domain knowledge to real-time recommend contingency signal plans that accommodate non-recurrent traffic, with the outputs from real-time traffic prediction at least 30 minutes in advance. Specifically, considering the rare occurrences of engagement of contingency signal plans for incidents, we propose to decompose the end-to-end recommendation task into two hierarchical models -- real-time traffic prediction and plan association. We learn the connections between the two models through metric learning, which reinforces partial-order preferences observed from historical signal engagement records. We demonstrate the effectiveness of our approach by testing this framework on the traffic network in Cranberry Township in 2019. Results show that our recommendation system has a precision score of 96.75\% and recall of 87.5\% on the testing plan, and make recommendation of an average of 22.5 minutes lead time ahead of Waze alerts. The results suggest that our framework is capable of giving traffic operators a significant time window to access the conditions and respond appropriately.
\end{abstract}


\maketitle

\section{Introduction}\label{sec:introduction}

Most traffic management centers (TMC) operate a coordinated signal system that relies on historically generated signal timings, coupled with real time technology, to manage day to day operations on the local network. Unfortunately, any planned or unplanned incidents (e.g. hazardous weather conditions, accidents, local events, etc.) on the network can cause catastrophic traffic gridlocks. According to FHWA in 2019, about half of congestion is nonrecurring, among which 25\% is caused by accidents, 15\% by weather and 10\% by work zones \cite{fhwa2019tmi}. To keep traffic flowing during these occurrences, local TMCs develop incident timing plan, or contingency traffic plan for their owned signalized intersections to manage incident-induced congestion. However, the effectiveness of existing incident management is often limited by late response time and excessive workload of traffic operators, and the main causes are two-fold: (i) the lack of real-time and advance awareness of road conditions. Traffic operators often react after receiving complaints, when gridlocks have affected local arteries for quite a long time; and (ii) the workload from verification of incidents and determination of signal plans. Traffic operators need to gather and analyze incident information (e.g. location, lane closure types, etc.) from multiple directives, such as cameras and travel information platforms. In other words, incident plans are determined with considerable manual efforts of integration and analysis on multi-source traffic data.

In this paper, we propose to improve incident management efficiency by introducing a decision-making framework that automates the data analysis process and learns to recommend signal plans even before official report of incidents. Specifically, instead of learning end-to-end mappings from road traffic states to action plans, we decompose the recommendation task into two subtask models in hierarchy -- traffic predictor and signal plan associator. For traffic prediction, we incorporate real-time data inputs monitored from crowdsourced Waze alerts and traffic sensors to trigger predictions of traffic delays in the network. A novel neural sequential learning model using encoder-decoder architecture with attention mechanism \cite{luong2015effective} is developed for this task. For plan associator, to incorporate domain knowledge from developed incident timing plans, we encode every incident plan into a matrix of plan keys characterized by its incident triggering conditions, which are derived by transportation experts, and a normalizer converts traffic predictor outputs to queries. By defining various metric features for measuring the closeness between traffic query and plan keys, we propose to learn a linear kernel of metrics, which gives a higher ranking score for every historically-engaged query key pair than the irrelevant ones.  $L_1$-regularized rank logistic regression model (RankLR) is used for this task. It is found that our recommendation system shows a precision score of 96.75\% and recall of 87.5\% on the testing plan, and an average of 22.5 minutes lead time ahead of Waze alerts for making plan recommendations. The results suggest that our decision-making framework is capable of giving traffic operators significant time to access the condition and reacting appropriately.

\section{Related Work}\label{sec:related-works}

\subsection{Traffic Signal Timings for Incident Management}

Most TMCs respond to traffic incidents by placing variable message signs, closing lanes or forcing turnings. Recently, studies have examined the optimization of traffic signal timings as an active management tool for non-recurring congestion \cite{mao2019traffic,abudayyeh2018traffic,gordon2016non,blandin2019method} . Traffic assignment models \cite{qian2012system,pi2017stochastic,ma2017variance,qian2013hybrid}, equipped with behavior models that characterize traveler's behavior changes after incidents \cite{zhu2010traffic,danczyk2017unexpected} and prediction-correction models \cite{he2012modeling}, are often built to simulate the time-dependent diverted traffic flow under pre-defined incident scenarios. The signal timings, optimized for a given incident scenario, then favorite specified directional movements to minimize the induced congestion \cite{koonce2008traffic}. Our study, which is built upon developed incident timing plans, refines the decision making process or transition logic of signal patterns, by recommending optimal incident signal plans to traffic operators ahead of time. The closest work to ours is the work of Ban et al.~\cite{ban2016decision}, which considers real-time incidents, traffic volumes and weather data for activation of signal control, like us. However, they aim to determine if adaptive traffic control system (ATCS) should be activated -- while we recommend detailed signal timing plans to traffic operators. Their model learns to classify level-of-service (LOS) outcomes of signal control activation with equal amount of before-after experiment data. We instead train the model to replicate TMC's manual operation, of which records are readily available even in small townships. Most importantly, they do not consider ahead-of-time recommendation with traffic prediction, which is one of the central contributions of this work.

\subsection{Traffic Prediction}

Data-driven models have become popular approaches for real-time traffic prediction. Recent models are built with spatiotemporal traffic flows, traffic events and incidents and weather data \cite{yao2019real,yao2016prediction,qian2018user,yang2019understanding,min2011real} to trigger traffic volume or speed prediction 5 to 30 minutes ahead. Historical average \cite{smith1996multiple}, linear models such as autoregression \cite{min2011real} or Lasso \cite{yang2019understanding,zhang2018user,qian2018user}, local regression and nearest neighborhood methods \cite{qian2018user,zheng2016urban}, graphical models \cite{sun2006bayesian} and deep neural networks \cite{yao2019real,lv2014traffic,ma2017learning} are common modeling choices. Encoder-decoder recurrent neural network \cite{cho2014learning} is a popular deep learning architecture initially proposed for machine translation. Researchers have applied it for predicting traffic sequences \cite{ma2019trafficpredict,yang2019deep}. Attention mechanisms \cite{bahdanau2014neural,luong2015effective} are often embedded between encoder-decoder stacks to reduce the burden of compressing all observed information at each time step. Our study applies encoder-decoder Gated Recurrent Unit (GRU \cite{cho2014learning}) with bilinear attention mechanism \cite{bahdanau2014neural} for predicting spatial traffic time-series in target traffic network.

\subsection{Metric Learning}

Metric learning, whose goal is to find appropriate similarity measurements of data points, was initially proposed for recommendation system, such as search engines, to customize rankings with user clickthrough logs \cite{chechik2010large,joachims2002optimizing,sculleylarge}. It has been adapted to zero-shot learning to classify instances of unseen classes during training \cite{akata2015label,akata2015evaluation,frome2013devise,romera2015embarrassingly,xian2016latent,socher2013zero}. Their approach is to project inputs and class attributes into the same feature space and associate them with a compatibility function with learned parameters. We employ metric learning to tackle the cold-start problem of signal plan recommendation. The lack of expert records is expected to exist during initial enabling of timing plans, or when new plans are added for expansion of signalized intersections. \\

\subsection{Our contribution}

 Our study can be differentiated from prior work in three ways:
\begin{enumerate}
\item We shorten the incident response time of traffic operators by combining approaches from both traffic prediction and recommendation systems. Other work either predicts traffic without prescribing actions, or determines signal timings with current road conditions.
\item  A novel hierarchical model is proposed which learns to recommend incident signal timings to traffic operators with domain knowledge and very few historical demonstrations.
\item Our model processes crowdsourced data in real-time for traffic prediction and incident management. Few existing works present data processing and feature engineering methods for Waze data feeds.
\end{enumerate}

\section{Dataset and Preprocessing}\label{sec:dataset}

\begin{figure}[h]
\centering
\includegraphics[width=0.85\linewidth]{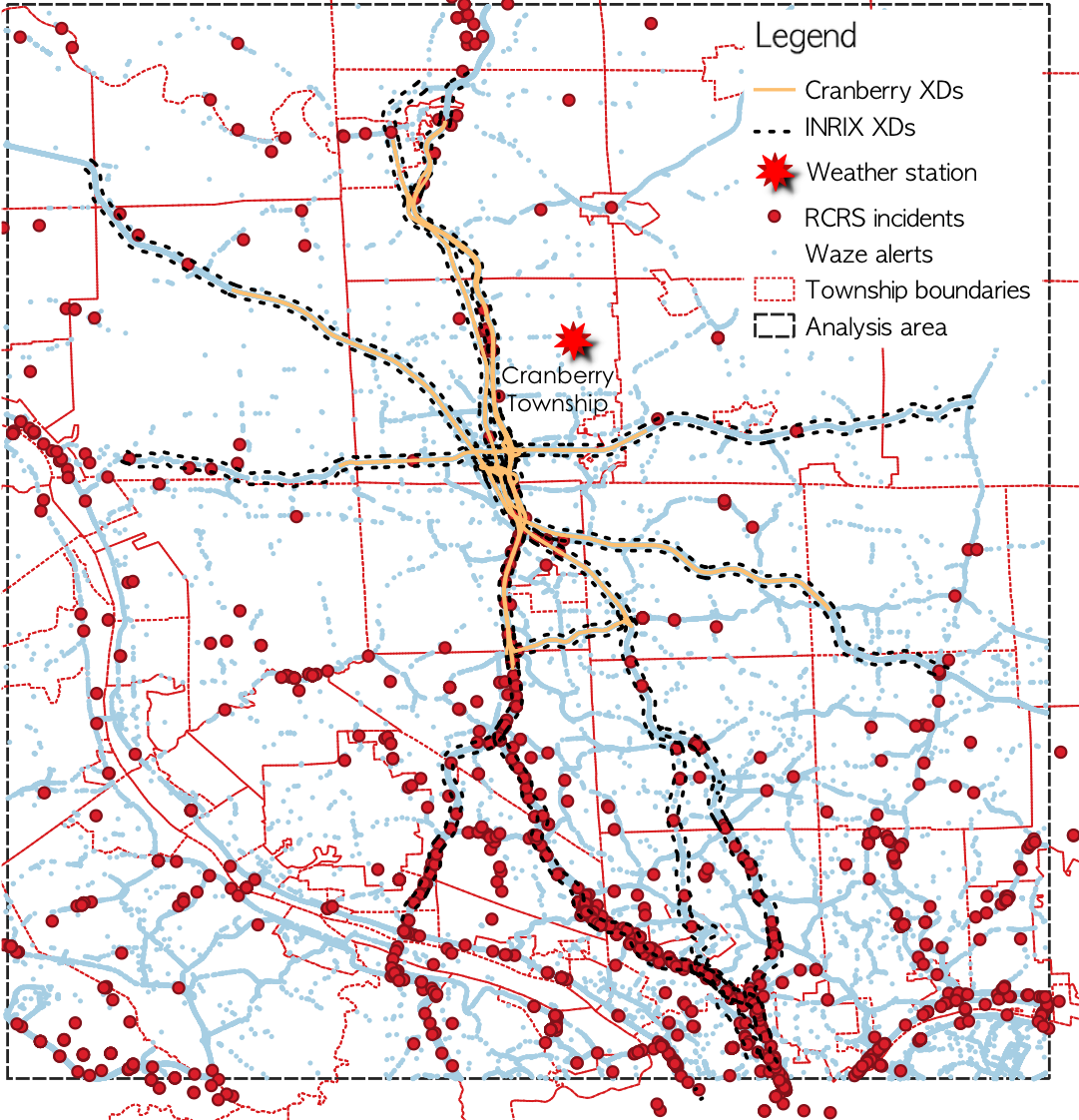}
\caption{Data sources used in this paper. \label{fig:data}}
\end{figure}

This section describes the data sources used in this work, which includes INRIX probed traffic speed data\footnote[1]{http://inrix.com/products/ai-traffic/}, PennDOT Road Condition Reporting System (RCRS) incident report\footnote[2]{https://www.penndot.gov/Doing-Business/OnlineServices/Pages/Developer-Resources.aspx}, Waze alerts\footnote[3]{https://www.waze.com/} and Weather Underground\footnote[4]{https://www.wunderground.com/}. \figref{fig:data} illustrates the data sources and collection area for Cranberry Township in Pennsylvania.

\subsection{INRIX Traffic Speed}

The INRIX traffic data were reported every 5 minutes for road segments georeferenced by INRIX XD code. Each data record includes the XD segment code, time stamp, observed speed (mph), average speed (mph), reference speed (mph) and two parameters for the confidence of the speed, namely confidence score and confidence value. We downloaded INRIX data between Jan 1, 2017 and Jul 21, 2019. We selected all XD segments in Cranberry Township, and Interstate freeways XDs within 30 minutes driving in our study. The selected XD segments are shown in \figref{fig:data}. Missing values are imputed with the last speed observations of this segment.

\subsection{Traffic Incident}

\subsubsection{PennDOT RCRS Incident Report}

RCRS data feeds, shown as red dots in \figref{fig:data}, provide real-time information for traffic incidents, roadwork, winter road conditions, and other events which cover all state-owned roads in 511PA road network. An incident record includes the incident location, road closure types, causes, and close and open time stamps. RCRS reports between 2017 and 2019 were provided by PennDOT. The incident records without location coordinates, and those lasting for more than 24 hours, are removed from the dataset.

\subsubsection{Waze Alerts}

Waze is a mobile navigation application that enables editing of the map with crowdsourced user reports. Users can report traffic crashes, congestion, hazards, or police traps on the road \cite{amin2018evaluating}. Waze data (blue dots) from Feb 9, 2019 to Jul 23, 2019 were collected from Waze GeoRSS API. We only consider incidents reported under accident or jam categories to remove most of the false alarms. However, we do not screen waze regarding the reliability score to lower our dependency on the external system.

\subsection{Weather Underground}

Weather Underground reports hourly weather measurements. Each entry contains temperature, pressure, dew point, humidity, wind speed, precipitation, pavement condition, and visibility, etc. The position of the chosen weather sensor is shown in \figref{fig:data}. The weather time series are resampled every hour and missing values are imputed with linear interpolation.

\section{Method}\label{sec:method}

This section first describes the data processing steps. We then present our model architecture that learns to recommend incident timing plans by two decomposed learnable models, namely, traffic predictor and signal plan associator.

\subsection{Data Processing}

 The proposed data processing pipeline integrates and transforms multi-source traffic speed, incident, weather and temporal data into representative features for sub-task models. We apply one-hot encoding for categorical variables. All processed features are scaled by Min-Max normalization.

\subsubsection{Speed Processing}\label{sec:spd-process}
 Two segment-level features, travel time index ($TTI_{itd}$) and slowdown speed ($SD_{itd}$), are extracted from raw traffic speed data to describe road conditions. We use $v_{itd}$ to denote the observed speed on XD $i$ at time $t$ on day $d$. To measure congestion on this segment, we use travel time index (TTI), which is defined as real-time travel
time divided by free-flow travel time, and can be computed by \equref{eq:tti}. To determine the reference (free-flow) speed $v_i^{ref}$ of an XD $i$, the 85 percentile of observed speed on that segment for all time periods (\equref{eq:ref-spd}) is used, which is the recommended approach for computing reference speed from probe-based speed data \citep{jha2018estimating}. A large value of $TTI_{itd}$ indicates the segment is congested.

\begin{gather}
TTI_{it}^d = \max (v_i^{ref}/v_{itd}, 1) \label{eq:tti}\\
v_i^{ref} = \mathbb{P}_{0.85}(v_{itd})\label{eq:ref-spd}
\end{gather}

 To encode flow spillbacks in the network, we propose slowdown speed ($SD_{itd}$). Slowdown speed, defined in \equref{eq:sd}, is computed by subtracting speed  $v_i$ from the mean speeds of $N_i$ upstream XDs of $i$, denoted as $\Gamma^{-1}(i)$.  A large value of $SD_{itd}$ indicates that back-of-queue slowdowns exist on segment $i$ and may infer the occurrence of traffic incident.

\begin{equation}
SD_{itd}  = \max[\frac{\sum_{j \in \Gamma^{-1}(i)} v_{jtd}}{N_i} - v_{itd}, 0] \label{eq:sd}\\
\end{equation}

\subsubsection{Incident Processing}
 We propose an integration and processing method for multi-source incident data. It deserves notice that Waze differs from RCRS that (i) Waze alerts are reported by road users, which often appears immediately after the occurrence of incidents, while RCRS documents TMC's road closures in response to incidents, which are inputted after actions are taken; (ii) Waze contains geographically point features indicating the position of road users, while RCRS is usually line feature for  the begin/end locations of road closure events; (iii) Waze has duplicate records for one incident, while RCRS is usually unique. Naturally, we assume that Waze and RCRS represent different status of a traffic incident: (i) road users first report incident occurrence on Waze; (ii) if the induced congestion calls for road closure,  TMC then takes actions, such as placing barricades on the affected road, and documents it on RCRS; (iii) traffic goes back to normal (incident clear), and Waze and RCRS are removed from the feed.
 
\begin{figure*}[ht]
\includegraphics[width=0.85\textwidth]{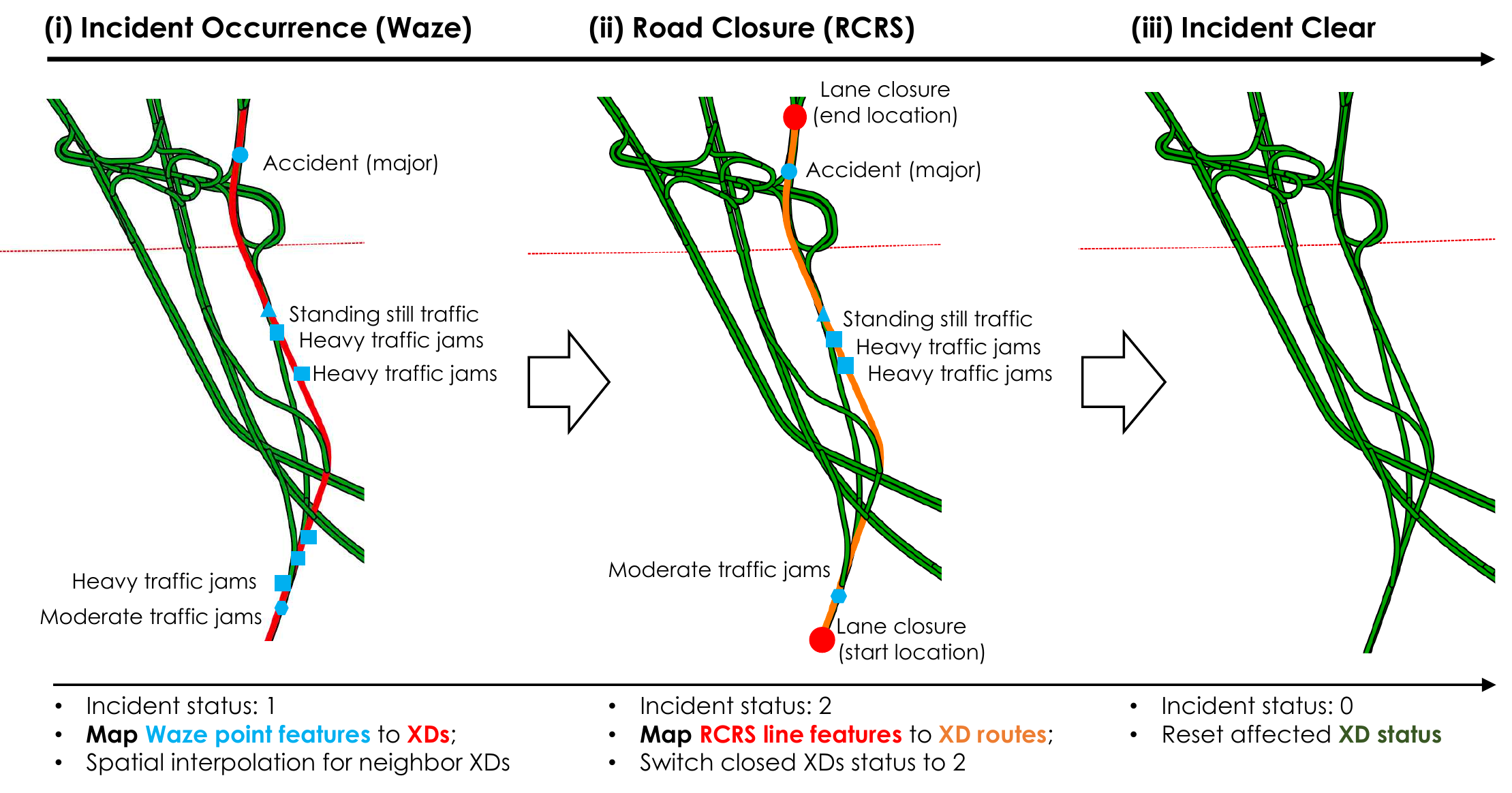}
\caption{Workflow for integration of multi-source incidents. \label{fig:incident}}
\end{figure*}

We thus develop the workflow in \figref{fig:incident} to integrate multi-source incident information by their location and status. For incident location, we first map incidents to their affected road segments. A vector $\mathbf{x}_{td}^{inc}=[x_{1td}^{inc},\hdots,x_{ntd}^{inc}]$, where $n$ is the number of segments in the network, is used to describe the spatial incident location at $t$. Note that for processing Waze point features, an interpolation step is performed at each time step to fill spatial gaps between affected segments. We add the middle segment along the route to the incident affected area if the shortest graph distance between both ends of two Waze alerts is just one neighbor. For incident status, we encode each element $x_{itd}^{inc} \in \mathbf{x}_{td}^{inc}$ as a three-category variable, where 0 is normal traffic, 1 means the incident is reported by Waze, and 2 denotes the road closures reported by RCRS. A max-gate operator then integrates multi-source incidents status on segments at each time step, i.e.,  the final value of $x_{itd}^{inc}$ is the largest status value mapped to segment $i$ by multi-source data. The proposed incident processing workflow is expected to represent a major incident as 1-2-0 or 1-2-1-0 along the time dimension, as shown in  \figref{fig:incident}.

\subsubsection{Weather and Time Features}
Weather features used in this paper include six continuous variables -- temperature, humidity, wind speed, pressure, visibility and hourly precipitation, and a binary variable -- pavement condition. Time features include five categorical variables: time-of-day, week-of-year, month-of-year, day-of-week, and holiday. For the cyclic month, week-of-year and time-of-day categorical variables, we use sine and cosine functions to transform them into a two-dimension vector $[t^{(sin)}_i, t^{(cos)}_i ]$:
\begin{gather}
t^{(sin)}_i = \sin(2\pi i/T)\\
t^{(cos)}_i = \cos(2\pi i/T)
\end{gather}
where $i$ denotes the week/month/time index and $T$ denotes the total weeks/months/time steps in a cycle. An advantage of this ``clockwise'' encoding is that each variable is mapped onto a circle such that the lowest value for that variable appears right next to the largest value (e.g. Jan is right to Dec). For day-of-week and holiday variables, we apply one-hot encoding after combining similar time features. Specifically, while Monday and Friday are encoded separately, Tuesday to Thursday are merged into one variable, so are Saturday, Sunday and official holidays.

\subsection{Model Architecture}

 The proposed incident plan recommender consists of two interconnected learnable models: traffic predictor and plan associator. The traffic predictor is an encoder-decoder recurrent neural network with attention mechanism \cite{cho2014learning}, which takes the speed (and slowdowns), traffic incidents, weather and time features and triggers  predictions of traffic speed time-series on target segments for the future 30 minutes. To incorporate domain knowledge of the developed incident plans, we obtain the signal timing manual from Cranberry Township Traffic Management Center and translate the plan triggering conditions into a matrix of plan attributes (keys). The plan associator then generates traffic queries from current and 30-min predicted future speed series and their closeness with plan keys are evaluated with self-defined metrics. The plan associator learns to rank incident plans by fitting a linear kernel of the proposed metrics from historical engagement records. The module is named as metric kernel in the remainder of this paper.

\subsubsection{Traffic Predictor}
 Predicting traffic beyond 5-10 minutes ahead is hard for traditional autoregressive time-series methods. Traffic on a road segment can change drastically due to traffic incidents, weather hazards or atypical traffic patterns in its proximity. In these cases, past traffic dynamics on the target road segments may have little useful information implying their future traffic states. A widely-used solution found in literature \cite{ermagun2018spatiotemporal,yang2019understanding,min2011real} is to take into account spatiotemporal correlations between target road segments and nearby segments. As it takes time for traffic to propagate, abnormal traffic nearby can work as longer-term predictors. In this paper, we model the traffic prediction problem as a sequence-to-sequence task \cite{bahdanau2014neural}, and the state-of-the-art architecture, encoder-decoder neural networks with attention mechanism \cite{luong2015effective}, is built to predict the future traffic flow sequence up to 30 minutes, with a resolution of 5 minutes.

\begin{figure*}[ht]
\includegraphics[width=0.85\textwidth]{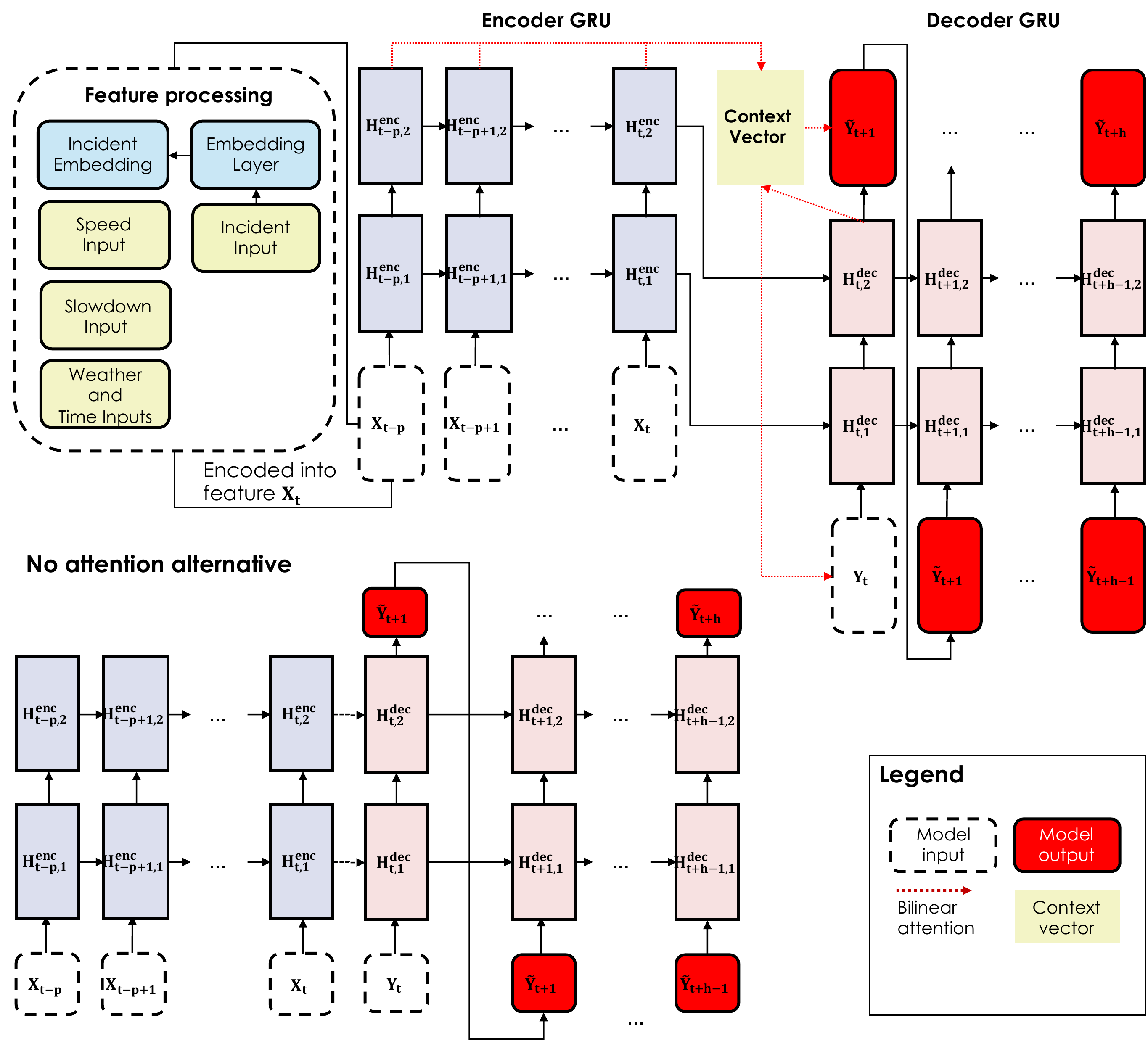}
\caption{Traffic predictor architecture. \label{fig:predictor}}
\end{figure*}

As shown in \figref{fig:predictor}, the architecture is comprised of an encoder model, a decoder model, and an attention model which queries the encoder dynamically via a context vector. Both encoder and decoder models use Gated Recurrent Unit \cite{cho2014learning}, which processes the feature input $\mathbf{X}_{t}$ sequentially into encoder hidden states $\mathbf{H}^{\text{enc}}_t$ and predicts the speed time-series $\tilde{\mathbf{Y}}_{t+h}$ on all target segments. Denote $h$ as the forecasting horizon, which ranges from 5 to 30 minutes. The definition of hidden state $\mathbf{H}^{\text{enc}}_t$ is consistent with original paper \cite{cho2014learning}. The encoder hidden states at the last time step $\mathbf{H}^{\text{enc}}_t$ is set as the initialized hidden states of decoder model. For decoding, the current speed on target segments $\mathbf{Y}_t$ is used as initial inputs to the model, and the predicted speed $\tilde{\mathbf{Y}}_{t+1}$ are fed as inputs for next step predictions in an autoregressive way.

In the standard encoder-decoder models, the encoder model attempts to compress all the observed information at each time step into an intermediate representation of fixed size. One way to address this issue is via an attention mechanism, where we keep references to the hidden states of the topmost encoder and query them dynamically for decoding. There are also some physical intuitions for applying attention mechanism. For short-term traffic prediction, we expect that traffic conditions on downstream segments close to the target segment are good indicators. However, for longer-term prediction, traffic conditions on downstream segments of variable distances (usually far away from target segments), which depend on current and predicted network traffic conditions, should be dynamically attended to, as congestion takes time to propagate backwards. Hence, a bi-linear attention mechanism proposed by \cite{bahdanau2014neural} is used in this paper. At each time step, the decoder computes the attention weights for each encoder output $\mathbf{H}^{\text{enc}}_t$ by a bi-linear correlation kernel with learned weights.  The context vector $c_t$ in \figref{fig:predictor} is the average of encoder output at each time step weighted by attention weights and is integrated with decoder hidden states to trigger the speed predictions.

\subsubsection{Plan Associator}

\begin{figure*}[ht]
\includegraphics[width=0.85\textwidth]{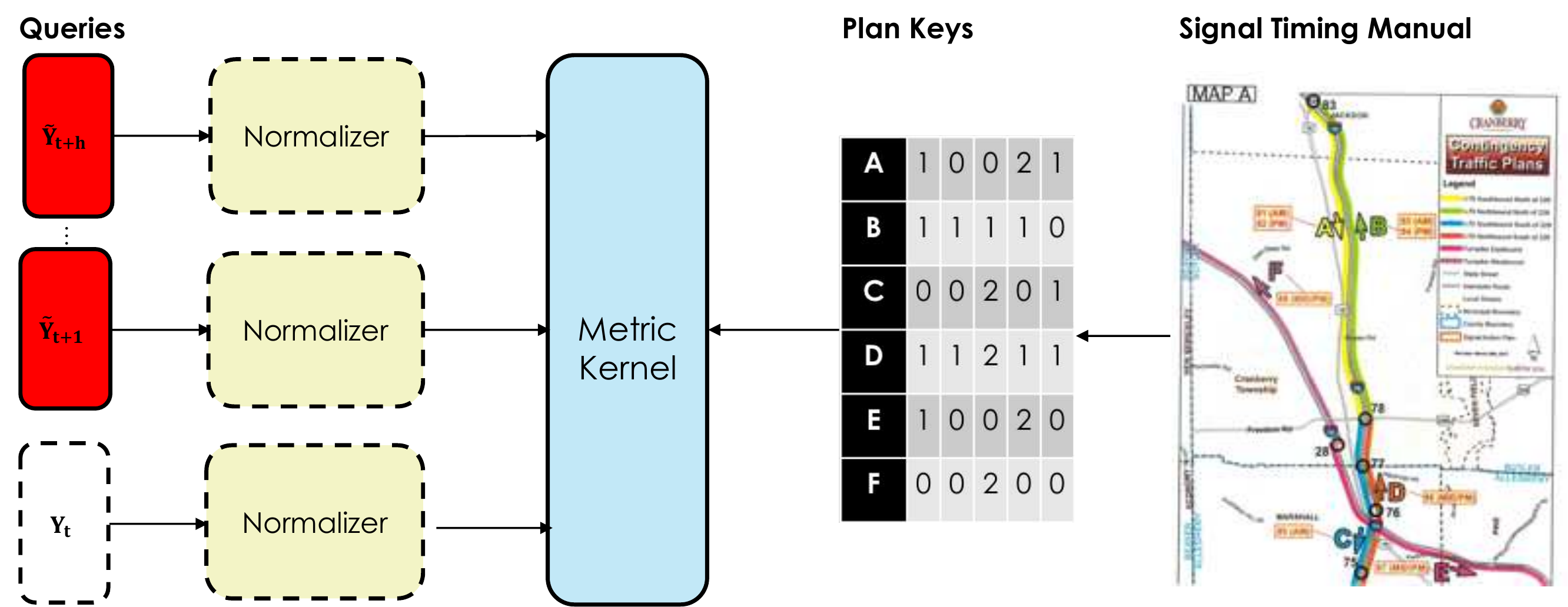}
\caption{Plan associator architecture. \label{fig:associator}}
\end{figure*}

 The plan associator learns to select from pre-defined decision-making rules, and combine them to recommend signal timing plans based on current and predicted traffic conditions on target network. This paper proposes to learn similarity metric kernel between network traffic queries, which are transformed from current and predicted speed time-series on target segments, and keys, which are the triggering conditions of each incident plan. As shown in \figref{fig:associator}, the plan associator architecture comprises of the encoding schemes of incident plan keys, the normalizer for network traffic prediction queries and the learned metric kernel.
 
\paragraph{Incident Signal Timing Plan}
 We obtained the signal timing manual from Traffic Management Center (TMC) in Cranberry Township. Cranberry Township is located at the junction of Interstates 79 and 76 (PA Turnpike). The township owns 18 active and 6 proposed signalized intersections on local arterial roads US 19, Freedom Road (3020) and Route 228. Six coordinated signal timing plans for different incident scenarios, which involves the control of the 18 active signals, have been developed by the TMC and used in this study. These incident signal timing plans are prepared by (1) simulating the diverted traffic flow under pre-defined incidents using PTV Visum\footnote[1]{http://vision-traffic.ptvgroup.com/en-us/products/ptv-visum/}, and (2) optimizing signal timing function and coordination by Synchro Studio\footnote[2]{https://www.trafficware.com/synchro.html}, with the diverted travel demand as inputs. The developed plans are as follows.

\begin{itemize}
\item \textit{Plan A} is an incident plan for managing incidents on I-79 Southbound south of Exit 83 (Zelienople / Jackson Township SR 528) and North of Route 228. Traffic signal timings were developed to favor US 19 southbound, from Old Route 19 / Victory Church Dr to Thorn Hill Rd. One entire network coordination zone with 175 second cycle lengths for both AM and PM peaks is activated;
\item \textit{Plan B} manages incidents on  I-79 Northbound and North of Route 228.  One entire network coordination zone, with half-cycling where possible is activated to favor US 19 northbound, from Emeryville Rd / Freeport Rd to Old Route 19 / Victory Church Dr. AM Peak cycle length is 180 seconds (90 seconds) and PM Peak cycle length is 210 seconds (105 seconds);
\item \textit{Plan C} is for incidents on I-79 Southbound and South of Route 228. One entire network coordination zone half cycling is activated to favor US 19 southbound. Cycle lengths are 200 seconds (100 seconds) for both peaks;
\item \textit{Plan D} is for incidents on I-79 Northbound and South of Route 228. One entire network coordination zone with 180 seconds cycle length is activated to favor movement on US 19 northbound;
\item \textit{Plan E} is for incidents on I-76 (PA Turnpike) Eastbound and East of Cranberry Township. The majority of the traffic was assumed to be heading east on SR 228 towards SR 8 back towards I-76 (PA Turnpike). One entire network coordination zone with half cycling (180/90 seconds) at the Turnpike Ramps and I-79 Ramps intersections is activated for PM peaks to favor this movement;
\item \textit{Plan F} is for incidents on I-76 Westbound and West of Cranberry Township. The majority of the traffic was assumed to be heading west on Freedom Rd towards SR 65. One entire network coordination zone with half cycling (180/90 seconds) at the Turnpike Ramps and I-79 Ramps intersections is activated to favor this movement during PM peaks.
\end{itemize}

\paragraph{Incident Plan Keys}

We process the incident plan triggering conditions presented above into plan key matrix. Traffic incidents (e.g. significant congestion or road closures) occurring on I-79 and I-76 are likely to spill back to the local network and causes catastrophic traffic gridlocks. 
 
\begin{figure}[h]
\includegraphics[width=0.85\linewidth]{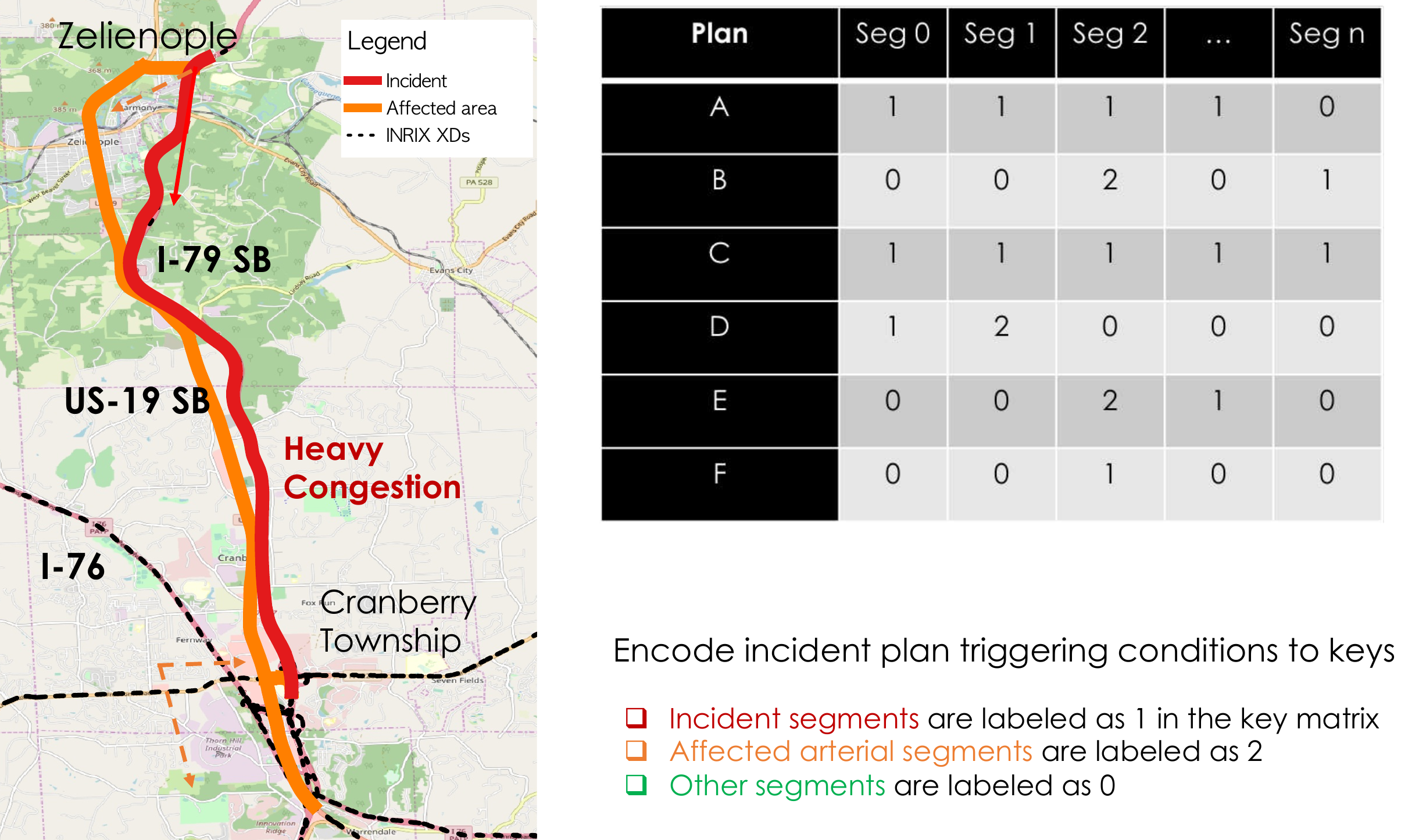}
\caption{Encoding scheme of incident plan keys. \label{fig:plan}}
\end{figure}

 For example, if a major incident occurs on I-79 SB north of Rt 228 causing severe congestion (Plan A), as shown in \figref{fig:plan}, traffic is expected to enter southbound US 19 at Zelienople and back up at Rt 228. A developed coordinated signal pattern A, which favor US 19 southbound movement, will be engaged.  To incorporate the domain knowledge from developed incident plans, we build a plan key matrix $\mathbf{P}$ to characterize the plan attributes and triggering conditions. As illustrated in \figref{fig:plan}, the entries of incident segment columns in $\mathbf{P}$ is set as 1 and affected arterial segment columns set as 2. A null plan is added to the matrix, with all columns set to 0, to represent that no plan is to implement.
 
\paragraph{Normalizer for Network Traffic Queries}
 We transform the speed outputs of traffic predictor to travel time index (TTI) by  \equref{eq:tti}. This intermediate step transforms traffic predictor outputs into a same scale for every segment.

\paragraph{Metric Learning}
 To learn a metric kernel that associates network traffic queries to plan key matrix, we define several metrics beforehand to characterize the closeness between traffic queries and plan keys, and fit a linear function of these metrics, which reinforces partial-order preference observed in engagement records. The metric kernel learns to select from pre-defined metrics, or decision rules, and combine them to determine signal timing plans. Three groups of metrics are defined to evaluate the relevance between network traffic queries at each time step and plan keys, which include:

\begin{enumerate}
\item \textit{Triggering precision:} We pre-define a threshold set $\{TTI^{\text{thres}}\}$ to detect if the query items exceed the threshold. The incident key matrix is also binarized to 0-1. For incident plans A-F, if any item of incident segments in queries is 1, the metric outputs 1 and otherwise 0. For null plan, the precision score between binarized key matrix and thresholded query vector are evaluated (0 is positive label), because the triggering segments of null plan are the whole target network. TTI thresholds, including 1.6, 2, 2.5, 5 and 10 are used.
\item \textit{Rule:} After thresholding and binarizing the query, \textbf{if} the overlapped terms between query and binarized key vector contain both incident and affected arterial segments, \textbf{then} their relevance is 1 and \textbf{otherwise} 0. The same set of thresholds are used.
\item \textit{Similarity:} The query vector is first upper-clipped by the threshold and scaled to 0-1 by min-max normalization. The inverse of euclidian distance between binarized key matrix and normalized query matrix is evaluated. The same set of thresholds are used.
\end{enumerate}

The derived 15 metrics between a network traffic query $q_k$ and plan key $p_k$ are evaluated for 7 (6 predicted and 1 current) time stamps in traffic queries and concatenated into $\mathbf{x}(q_k, p_i) \in \R^{105}$. Rank Logistic Regression (RankLR) proposed in \cite{sculleylarge} is applied to fit a linear kernel of the developed metrics. We build a dataset $D$ containing all network traffic query and plan pairs in our records, with the engaged pairs in  $D_+^{(k)}$ and irrelevant ones in $D_{-}^{(k)}$. We create a pairwise dataset by drawing any pair $i$ from $D_+^{(k)}$ and one $j$ from $D_{-}^{(k)}$, evaluating their respective metric vector $ \mathbf{x}(q_k, p_i)$ and $\mathbf{x}(q_k, p_j)$, and computing the difference $\mathbf{x}_{ij}^p = \mathbf{x}(q_k, p_i) - \mathbf{x}(q_k, p_j)$. These $\mathbf{x}_{ij}^p$ are set as positive samples for pairwise learning. A reverse operation is also conducted by choosing one pair from $D_{-}^{(k)}$ and one from $D_+^{(k)}$, and these are set as negative samples. $L_1$-regularized logistic regression is applied on the pairwise dataset to fit a linear kernel of the developed metrics so that the log loss in \equref{eq:rank-lr} is minimized. The algorithm finds  $\mathbf{w}$ that gives a higher ranking score for every relevant query-key pair than the irrelevant ones.

\begin{equation}\label{eq:rank-lr}
\min_{\mathbf{w}}  \frac 1 P \sum_{i=1}^P \big[y^{p}\log(\mathbf{w}^T \mathbf{x}^{p}_{ij}) + (1-y^{p})\log(1-\mathbf{w}^T\mathbf{x}^{p}_{ij}) \big]+ C \|\mathbf{w}\|_1
\end{equation}

\paragraph{Determination of Incident Signal Plan}

Recommendation score $s_{ij}$ of a query-key pair is defined in \equref{eq:score} as the linear combination of the defined metric evaluations, weighted by the learned $\mathbf{w}$. The incident signal plan with the highest recommendation score is activated or transitioned to. For transition between signal timing plans, we've added a trigger of at least 20 minutes between pattern changes as the system takes time to react, and traffic control can be significantly less efficient during the transition. Note that the incident signal timings are turned off if null plan is activated.

\begin{equation}\label{eq:score}
s_{ij} = \mathbf{w}^T\mathbf{x}(q_k, p_j)
\end{equation}

\subsubsection{Hyperparameters and Training}

The encoder-decoder neural network is implemented in PyTorch. For all GRUs, we use: Tanh activation, 256 hidden dimensions, 2 layers and recurrent dropout of 0.2. The embedding layer for incidents has 3 dimensions. Attention has 256 hidden states. We train the network using Adam optimizer for a maximum of 200 epochs and early-stops if validation error does not decrease for 5 epochs. Learning rate of 0.0005, teacher-forcing ratio of 0.5 and mini-batch size of 32 are used. For RankLR, we use an $L_1$ penalty of $C=1$ for all models.

\subsubsection{Baselines}
We experiment with four traffic prediction baselines.

\paragraph{Historical-average}
A baseline which uses day-of-week speed profiles averaged over the past one-month window as speed predictions for the future 30 minutes.

\paragraph{Latest-observation}
A baseline which uses the latest observed speed on the segment as speed predictions for the future 30 minutes.

\paragraph{LASSO}
Linear regression models with $L_1$ regularization, i.e., LASSO \cite{tibshirani1996regression,yang2019understanding}, which use the same feature set as the neural network model, are built for each segment $i$ and prediction horizon $h$ independently. The model learns the weights $w_h^{(i)}$ such that the loss in \textbf{Equation~\ref{eq:lasso}} is minimized. $L_1$ regularization helps the model select the most critical features that are linearly related to the response. $L_1$ penalty hyperparameter $\alpha_{ih}$ controls the number of selected features, and is tuned by cross validation on training set.

\begin{equation}\label{eq:lasso}
\min_{\mathbf{w}_h^{(i)}} \| \mathbf{y}_{t+h}^{(i)} - [\mathbf{X}_{t-p},...,\mathbf{X}_t] \mathbf{w}_h^{(i)}  \|_2^2 + \alpha_h^{(i)} \|\mathbf{w}_h^{(i)} \|_1
\end{equation}

\paragraph{GRU-no-attention}
 A non-linear GRU model baseline. This baseline removes the attention mechanism in our model. The same hyper-parameters of our model are used.\\

\begin{figure*}[ht]
\begin{minipage}[t]{0.85\textwidth}
  \includegraphics[width=\linewidth]{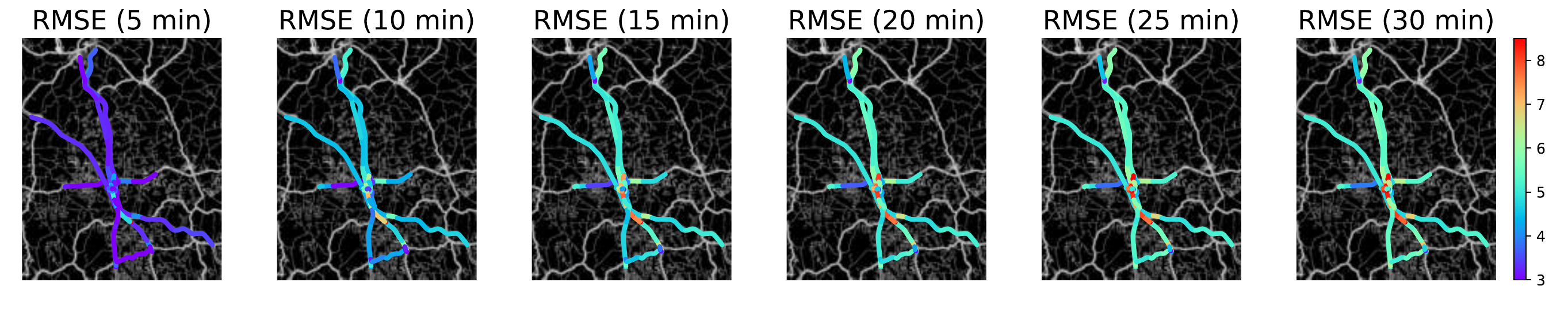}
  \subcaption{Speed prediction performance of \textbf{encoder-decoder-attention} with 5-30 minutes prediction horizon.}
  \label{fig:speed}
\end{minipage}
\hfill
\begin{minipage}[t]{0.85\textwidth}
  \includegraphics[width=\linewidth]{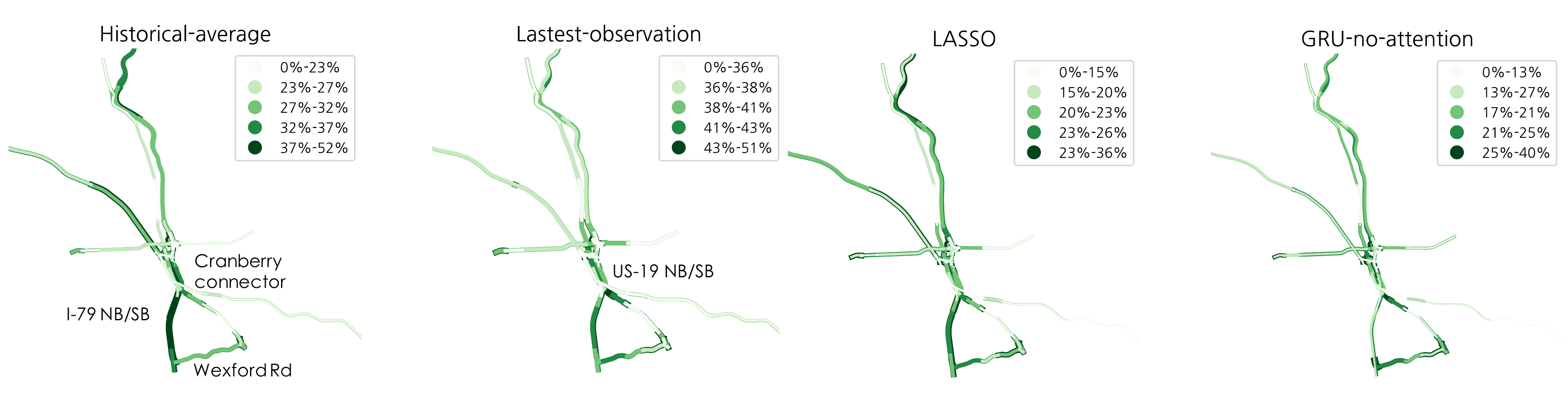}
  \subcaption{Percentage improvement of our model against other baselines for 30-min ahead speed prediction.}
  \label{fig:comparison}
\end{minipage}
\caption{Visualization of traffic speed prediction model performances. \label{fig:prediction}}
\end{figure*}

\section{Experiments}
The experiments were run on a Linux workstation with a P4000 GPU. Traffic and multi-source data from 5:30 AM to 8:55 PM are selected, which have in total 29,520 data samples. Four signal engagement demonstrations, one for each incident plan (A,C,D,F),  are available for training and evaluation. It deserves notice that we removed the whole four days' data, during which incident signal timings were engaged, for training the traffic predictor. The experiment setting ensures that ground-truth traffic conditions are not abused for prescribing incident plans. However, drawbacks are that traffic predictor only learns traffic dynamics when no human impact, i.e., manual changes of signal plans, is involved. In this study, we assume that delayed actions of traffic operators do not impact traffic dynamics much during catastrophic traffic gridlocks. Therefore, our proposed models are suitable in initial enabling phase of our decision-making framework, when late response of traffic operators still exists. However, when traffic operator's response time is largely reduced, new traffic predictor considering manual change of signal plans should be built.

For testing traffic prediction and incident plan recommendation performances, we adopt different evaluation methods:

\begin{itemize}

\item For traffic predictor, 80\% of the data samples are used for training and 20\% for testing. Hyper-parameters of LASSO are tuned by 5-fold cross-validation on the training samples. Other baselines do not require tuning. Root-Mean-Squared error (RMSE) and Mean-Absolute-Percentage-Error (MAPE) of speed predictions on the test set is computed for model comparison. Note that when computing MAPE, we deliberately divide the absolute error by the predicted values, so that negative errors (e.g., unable to predict congestion growth) are with a heavier penalty. Our MAPE is defined in \equref{eq:mape}:

\begin{equation}\label{eq:mape}
MAPE_t = \frac 1 n \sum_{i=1}^n \frac{|A_t - F_t|}{F_t}
\end{equation}
where $A_t$ is the actual value and $F_t$ is the forecast value.

\item For the recommender model, we first fit it with all engagement records to examine the model weights. Then, we apply leave-one-out evaluation, i.e., every three engagement records are used for training plan associator and the recommendation is evaluated on the remaining record. We compute macro precision and recall scores of our recommended plans for all time stamps during signal engagement periods, and visualize the whole recommender scoring behaviors for each testing plan.

\end{itemize}

\begin{table*}[ht]
\caption{RMSE prediction error (mph) for different prediction horizons on the test samples.}
\label{tab:results-traffic}
\begin{tabular}{@{}lllllll@{}}
\toprule
\textbf{Model} & 5 min & 10 min & 15 min & 20 min & 25 min & 30 min\\ \midrule
Encoder-decoder-attention (our model) & {\ul \textbf{3.187}} & {\ul \textbf{3.579}} & {\ul \textbf{3.461}} & {\ul \textbf{3.388}} & {\ul \textbf{3.417}} & {\ul \textbf{3.623}} \\
Historical-average & 5.306 & 5.306 & 5.306 & 5.306 & 5.306 & 5.306 \\
Latest-observation & 3.546 & 5.008 & 5.609 & 5.848 & 5.987 & 6.092 \\
LASSO & 3.297 & 4.207 & 4.460 & 4.557 & 4.606 & 4.647 \\
GRU-no-attention & 3.343 & 4.129 & 4.318 & 4.404 & 4.470 & 4.541 \\ \bottomrule
\end{tabular}
\end{table*}

\begin{table*}[h]
\caption{MAPE prediction error (\%) for different prediction horizons on the test samples.}
\label{tab:results-traffic-mape}
\begin{tabular}{@{}lllllll@{}}
\toprule
\textbf{Model} & 5 min & 10 min & 15 min & 20 min & 25 min & 30 min\\ \midrule
Encoder-decoder-attention (our model) & {8.39\%} & {\ul \textbf{8.43\%}} & {\ul \textbf{9.24\%}} & {\ul \textbf{9.48\%}} & {\ul \textbf{9.66\%}} & {\ul \textbf{9.78\%}} \\
Historical-average & 11.73\% & 11.73\% & 11.73\% & 11.73\% & 11.73\% & 11.73\% \\
Latest-observation & 9.05\% & 15.80\% & 18.92\% & 20.35\% & 21.22\% & 21.94\% \\
LASSO & {\ul \textbf{8.05}}\% & 13.43\% & 15.63\% & 16.02\% &  17.46\% & 17.52\% \\
GRU-no-attention & 8.13\% & 11.05\% & 11.48\% & 12.30\% & 13.09\% & 12.05\%\\ \bottomrule
\end{tabular}
\end{table*}

\begin{table*}[h]
\caption{Plan associator kernel weights (feature name: horizon-thres-metric).}
\label{tab:weights}
\begin{tabular}{@{}llllll@{}}
\toprule
\textbf{Feature} & \textbf{Weight} & \textbf{Feature} & \textbf{Weight} & \textbf{Feature} & \textbf{Weight} \\ \midrule
5min-2-similarity & 2.098 & 0min-1.6-rule & 0.321 & 0min-2.5-rule & 0.073 \\
10min-2-similarity & 2.07 & 5min-2.5-rule & 0.318 & 20min-2-rule & 0.071 \\
5min-2-precision & 1.459 & 15min-5-rule & 0.293 & 30min-2-similarity & 0.059 \\
25min-1.6-rule & 1.318 & 0min-2-rule & 0.29 & 25min-1.6-precision & 0.052 \\
10min-5-similarity & 1.062 & 10min-1.6-rule & 0.239 & 0min-10-similarity & 0.044 \\
25min-10-similarity & 0.746 & 5min-1.6-precision & 0.202 & 15min-2.5-rule & 0.04 \\
20min-10-similarity & 0.593 & 15min-2-rule & 0.197 & 20min-2-precision & 0.038 \\
25min-2.5-precision & 0.577 & 20min-5-rule & 0.177 & 15min-1.6-precision & 0.037 \\
15min-2-precision & 0.539 & 5min-1.6-rule & 0.17 & 20min-2.5-similarity & 0.036 \\
20min-1.6-rule & 0.535 & 30min-1.6-rule & 0.138 & 25min-2-rule & 0.033 \\
30min-2.5-rule & 0.425 & 5min-10-similarity & 0.124 & 25min-2.5-rule & 0.029 \\
25min-5-rule & 0.421 & 20min-2.5-precision & 0.118 & 30min-1.6-precision & 0.028 \\
30min-5-precision & 0.403 & 25min-5-similarity & 0.092 & 30min-2.5-precision & 0.015 \\
30min-2-rule & 0.375 & 5min-5-similarity & 0.083 & 0min-1.6-precision & 0.015 \\
15min-1.6-rule & 0.36 & 10min-2.5-rule & 0.081 & 10min-2-rule & 0.012 \\
10min-10-similarity & 0.339 & 15min-2.5-precision & 0.078 &  &  \\
0min-5-rule & 0.338 & 30min-10-distance & 0.078 &  &  \\ \bottomrule
\end{tabular}
\end{table*}

\subsection{Traffic Prediction}
Results of our model against other methods are presented in \tabref{tab:results-traffic} and \tabref{tab:results-traffic-mape} and visualized by segment in \figref{fig:prediction}. It is found that our encoder-decoder attention model outperforms speed prediction baselines for all prediction horizons within 30 minutes, and the performance improvement is more obvious for longer-term prediction and for predicting growth of traffic. When comparing the RMSE error, as shown in \tabref{tab:results-traffic}, for 5-minute ahead prediction, we find very less usage of incorporating multi-source data and applying complex model architecture, since \textbf{latest-observation} method is almost of the same performance as the best model. However, for longer-term prediction, such as those larger than 10 minutes, \textbf{latest-observation} performs even worse than \textbf{historical-average} which does not use any real-time data.  \textbf{GRU-no-attention} shows similar performances to \textbf{LASSO} on our dataset, but the model with attention mechanism added presents much better results for longer-term predictions. However, when compared with MAPE error in  \tabref{tab:results-traffic-mape}, our model outperforms other methods significantly. The results show that our model gains performance improvement mostly from reducing negative errors, i.e., better at predicting congestion growth.

To further locate the sources of performance improvement, we visualize the percentage RMSE improvement of our method against other baselines by road segment in \figref{fig:prediction}. It is found that most improvement comes from better predictions of traffic on I-79 NB/SB, Wexford Rd and US-19 NB, and Cranberry Connector. I-79 SB/NB south of Rt 228 are two main sources of traffic incidents in Cranberry Townships. A common incident-induced congestion pattern in Cranberry Township is that an incident on I-79 SB/NB closes part of the road, and traffic flows into US-19 as an alternative, hence routing to/from it via Wexford Rd. Our model learns to capture this pattern better than other methods. Note that this property is very useful for recommending proper incident plan ahead of time. However, if compared with \textbf{GRU-no-attention}, most improvement comes from better predictions on Cranberry Connector Ramp. This might suggest that \textbf{GRU-no-attention} captures the easier I-79 incident-induced traffic pattern well, but is hard to compress the traffic flow states in the complex Cranberry Connector without attention.

\begin{figure*}[ht]
\includegraphics[width=0.85\textwidth]{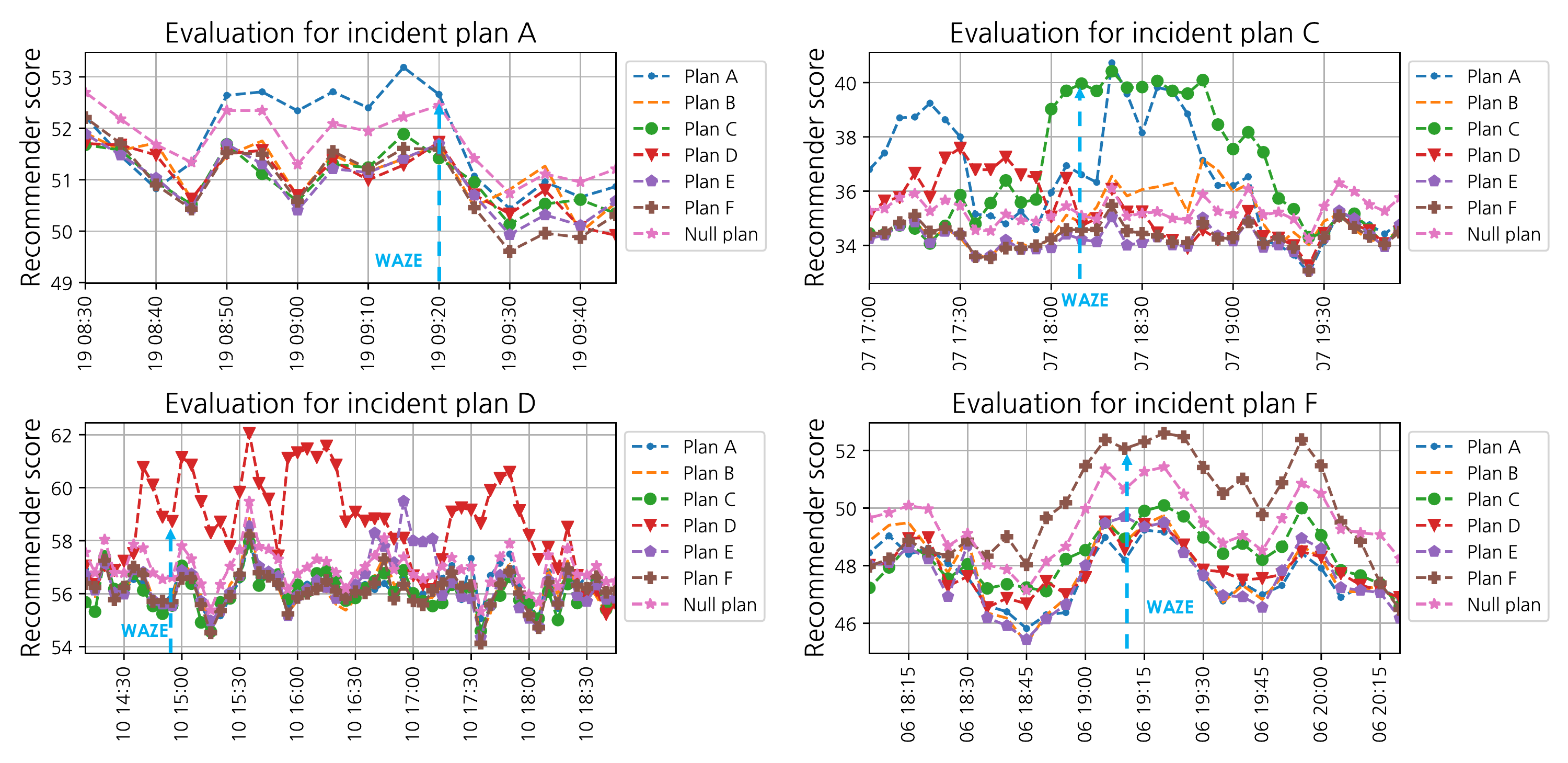}
\caption{Visualization of plan recommender model performances. \label{fig:recommend}}
\end{figure*}

\subsection{Incident Plan Recommendation}

We fit the metric kernel using all engagement records to examine the power of pre-defined metrics. The signal engagement periods, in total, span across 85 time stamps (425 minutes) and 41,334 $(85 \times 425)$ combinations of query-key evaluation pairs can thus be generated. The model selects 48 features from the 105 pre-defined evaluation metrics. Clearly, as shown in \tabref{tab:weights}, the selected features are reasonable. They are across all three types of metrics, different TTI thresholds and prediction horizons, as these features together are expected to stably determine the occurrences and clearance of congestion on plan triggering segments. In addition, all of the selected features have positive weights, indicating that the selected metrics alone are effective for measuring the association between network traffic queries and incident plan keys. Interestingly, it is found that instead of using current observations, most powerful metrics with larger weights are evaluated on the predicted traffic series across multiple horizons. This also meets our expectation, as a trend of traffic conditions are usually easier for decision making.

We then evaluate the generalization performances of plan recommender using leave-one-out strategy. Since each incident plan has only one engagement record, by using this evaluation method, the training and testing set become disjoint that unseen class labels during training appear in testing set. This issue is known as zero-shot learning problem. Although end-to-end classification methods cannot generalize to unseen classes, our method is expected to do it by learning an intermediate association between class attributes (incident plan keys) and features (network traffic queries) instead. In other words, the metric kernel learns to project network traffic queries into the space of plan keys.  As shown in \figref{fig:recommend}, our recommender triggers appropriate plans for all of the four test cases, and also for switching between null plan and incident plans. The recommendation precision and recall during signal engagement periods, if averaged over four test cases and all time stamps, are 96.75\% and 87.5\%. The inconsistency between the stopping  period of engagement causes the relatively low recall, where our recommender often stops the plan earlier than traffic operators. In addition, it is also found that if compared with the earliest incident report time from Waze in \figref{fig:recommend}, our model can trigger recommendations 10 minutes ahead for plan C, 35 minutes ahead for plan F, 30 minutes for plan A and 15 minutes for plan D. An average of 22.5 minutes advance recommendation performance can be achieved by our method.

\section{Conclusion}

This paper proposes a novel decision making framework which incorporates real-time data inputs monitored from crowdsourced Waze alerts and traffic sensors for traffic prediction, and constructs a learnable recommendation system for triggering incident signal plans ahead of time with the outputs from traffic predictor. The novelty of this work comes from decomposition of the end-to-end incident plan classification problem into two subtask models, i.e., a traffic predictor that outputs network traffic time-series for the future 30-minute horizon, and a plan associator, which transforms incident plan triggering conditions and predicted network traffic series into keys and queries with domain knowledge, and learns to associate them from historical signal plan engagement records.

The traffic prediction task is formulated as a sequential learning problem in this paper, and lagged spatiotemporal traffic speed, traffic incidents, weather and time features are embedded as source inputs for predicting traffic speed series on target segments.  We develop a new data processing pipeline for incorporating multi-source incident feeds from crowdsourced Waze alerts and PennDOT incident reports by their location and status, and encoding them into representative data features. A Gated-Recurrent-Unit neural network model using encoder-decoder architecture with bilinear attention mechanism is proposed for this task. Results show that our traffic prediction model outperforms other baselines, especially for longer-term traffic prediction and for predicting incident-induced congestion. The sources of prediction improvement are tracked and the model is found to be capable of capturing typical traffic patterns on our dataset. We incorporate domain knowledge from developed incident timing plans to constrain the model learning. An encoding scheme for transforming triggering conditions of each incident plan into the plan attribute keys is proposed and a normalizer for converting traffic predictor output to traffic queries is applied. By defining various metric features for measuring the closeness between traffic query and plans, we propose to learn a linear kernel of the proposed metrics, which gives a higher ranking score for every relevant query-key pair than the irrelevant ones from engagement records. Rank Logistic Regression (RankLR) model with $L_1$ penalty is used for this task. We find that the selected metric features are reasonable and are across all metric types, travel time index thresholds and prediction horizons. The recommendation system shows a precision score of 96.75\% and recall of 87.5\% on unseen testing plans. Our model can trigger an average of 22.5 minutes advance recommendation, in particular, 10 minutes ahead for plan C, 35 minutes ahead for plan F, 30 minutes for plan A ,and 15 minutes for plan D. Our proposed framework is expected to give traffic operators significant time to access the condition and react appropriately. In addition, our recommender has been shown to effectively recommend unseen plans in training. This generalization property makes our method an appropriate initializer for cold-start recommendation of new incident plans without engagement records, which may be created recently for expansion of signalized intersections.

\begin{acks}
This research is funded by Carnegie Mellon University’s Traffic21 Institute and Mobility21. Mobility 21 is a national University Transportation Center on mobility funded by U.S. Department of Transportation. The authors specially thank Jason Dailey, Kelly Maurer, and Marty McKinney from Cranberry Township Public Works for providing the contingency signal plan document and engagement records.
\end{acks}

\bibliographystyle{ACM-Reference-Format}
\bibliography{references}

\end{document}